\documentclass{aastex63}
\shorttitle{Hadronic vs leptonic models}
\shortauthors{Liu et al.}
\graphicspath{{./}{figures/}}

\begin{document}

\title{Hadronic vs leptonic models for $\gamma$-ray emission from VER J2227+608}

\correspondingauthor{Siming Liu}
\email{liusm@pmo.ac.cn, zhd@pmo.ac.cn}

\author[0000-0003-1039-9521]{Siming Liu}
\author{Houdun Zeng}
\author{Yuliang Xin}
\affiliation{Key Laboratory of Dark Matter and Space Astronomy, Purple Mountain Observatory, Chinese Academy of Sciences, Nanjing 210023, Peopleʼs Republic of China}
\affiliation{School of Astronomy and Space Science, University of Science and Technology of China, Hefei 230026, Peopleʼs Republic of China}
\author{Hui Zhu}
\affiliation{National Astronomical Observatories, Chinese Academy of Sciences, CAS, Jia-20 Datun Road, Chaoyang District, 100012 Beijing, PR China}



\begin{abstract}

Recent observations of VER J2227+608 reveal a broken power $\gamma$-ray spectrum with the spectral index increasing from $\sim 1.8$ in the GeV energy range to $\sim 2.3$ in the TeV range.
Such a spectral break can be attributed to radiative energy loss of energetic electrons in the leptonic scenario for the $\gamma$-ray emission, which, in combination with characteristic age of the nearby pulsar, can be used to constrain magnetic field in the emission region. We show that the radio and X-ray observations can also be explained in such a scenario. In the hadronic scenario, the spectral break can be attributed to diffusion of energetic ions in a turbulent medium and detailed spectral measurement can be used to constrain the diffusion coefficient. These two models, however, predict drastically different spectra above 100 TeV, which will be uncovered with future high-resolution observations, such as LHAASO and/or CTA.

\end{abstract}



\section{Introduction}
\label{intro}

Identifying PeVatron plays an essential role in addressing the origin of cosmic rays \citep{2020arXiv200513699A}. Although it is generally accepted that cosmic rays in the GeV and TeV energy ranges should be associated with particle acceleration by shocks in old and young supernova remnants (SNRs), respectively \citep{2017ApJ...844L...3Z}, the $\gamma$-ray spectra of SNRs never extend beyond 100 TeV \citep{2019ApJ...874...50Z}, implying that alternative high-energy sources are needed to account for cosmic rays in the PeV energy range \citep{2016Natur.531..476H}. Recently, HAWC discovered several sources beyond 50 TeV and 4 of their spectra extend beyond 100 TeV \citep{2019arXiv190908609H}. However, all these sources appear to be associated with powerful pulsars with evidence for spectral softening toward high energies.

VER J2227$+$608 was discovered by the VERITAS collaboration \citep{2009ApJ...703L...6A}, confirming earlier detection of TeV emission from this direction by the Milagro experiment \citep{2007ApJ...664L..91A}. Fermi observations revealed a GeV counter part with a hard spectrum with the best fit spectral index of $1.81\pm0.16$ \citep{2019ApJ...885..162X}. Recent HAWC observations of this source, in combination with VERITAS measurement, uncovered a single power-law spectrum in the TeV energy range with a spectral index of $\sim 2.3$ \citep{2020arXiv200513699A}. In the hadronic scenario for the $\gamma$-ray emission, the authors obtained a lower limit of 800 TeV for the cutoff energy of the parent proton distribution, making it a promising PeVatron. Their observations, however, do not exclude the possibility of spectral softening beyond 100 TeV. A leptonic scenario may also fit their data.

In this letter, we carry out detailed modeling of multi-wavelength emission from VER J2227$+$608 and the associated pulsar wind nebula PSR J2229$+$6114 and SNR G106.3$+$2.7 \citep{2000AJ....120.3218P,2001ApJ...547..323H, 2001ApJ...560..236K}. \footnote{Throughout this paper, we will not draw distinction between SNR G106.3+2.7 and VER J2227+608 unless specified otherwise.} The related observations are summarized in section \ref{obs}. In section \ref{Mod}, we present the key results of our spectral modeling. Conclusions are drawn in section \ref{con}.

\section{Observations}
\label{obs}

\begin{figure}[!htpb]
\centerline{\includegraphics[width=0.3\textwidth, angle=0]{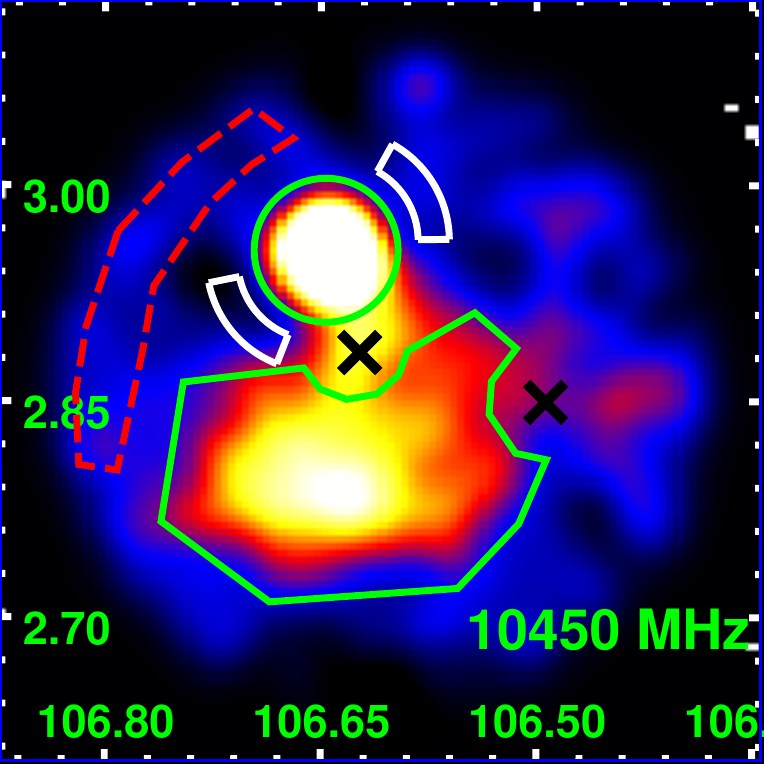}\includegraphics[width=0.3\textwidth, angle=0]{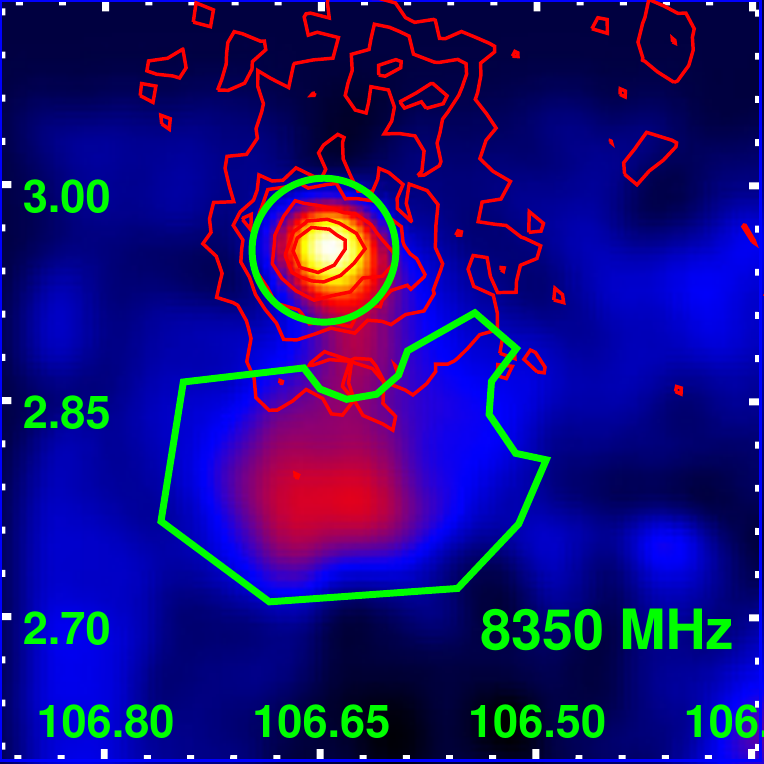}\includegraphics[width=0.3\textwidth, angle=0]{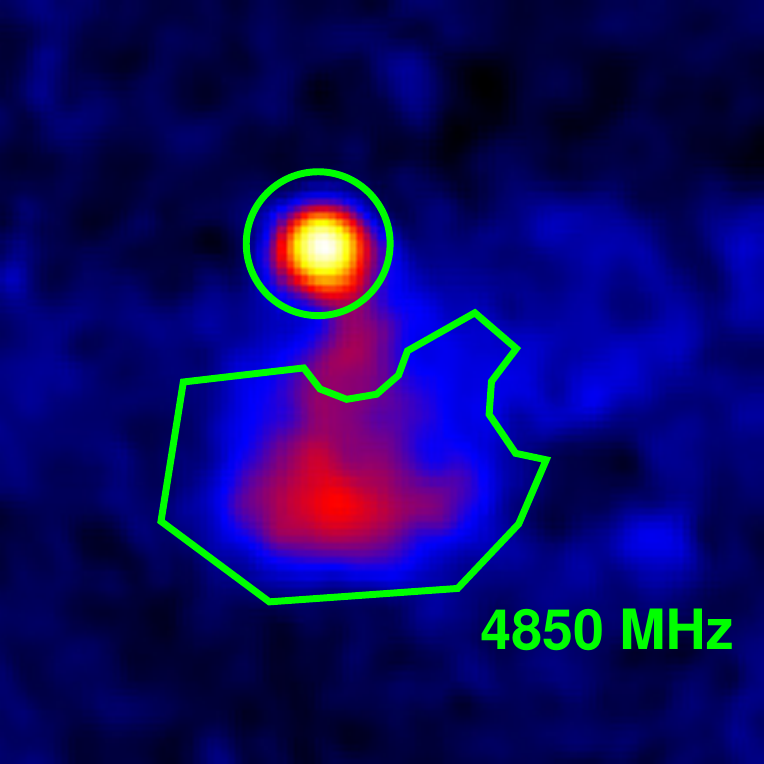}}
\centerline{\includegraphics[width=0.6\textwidth, angle=0]{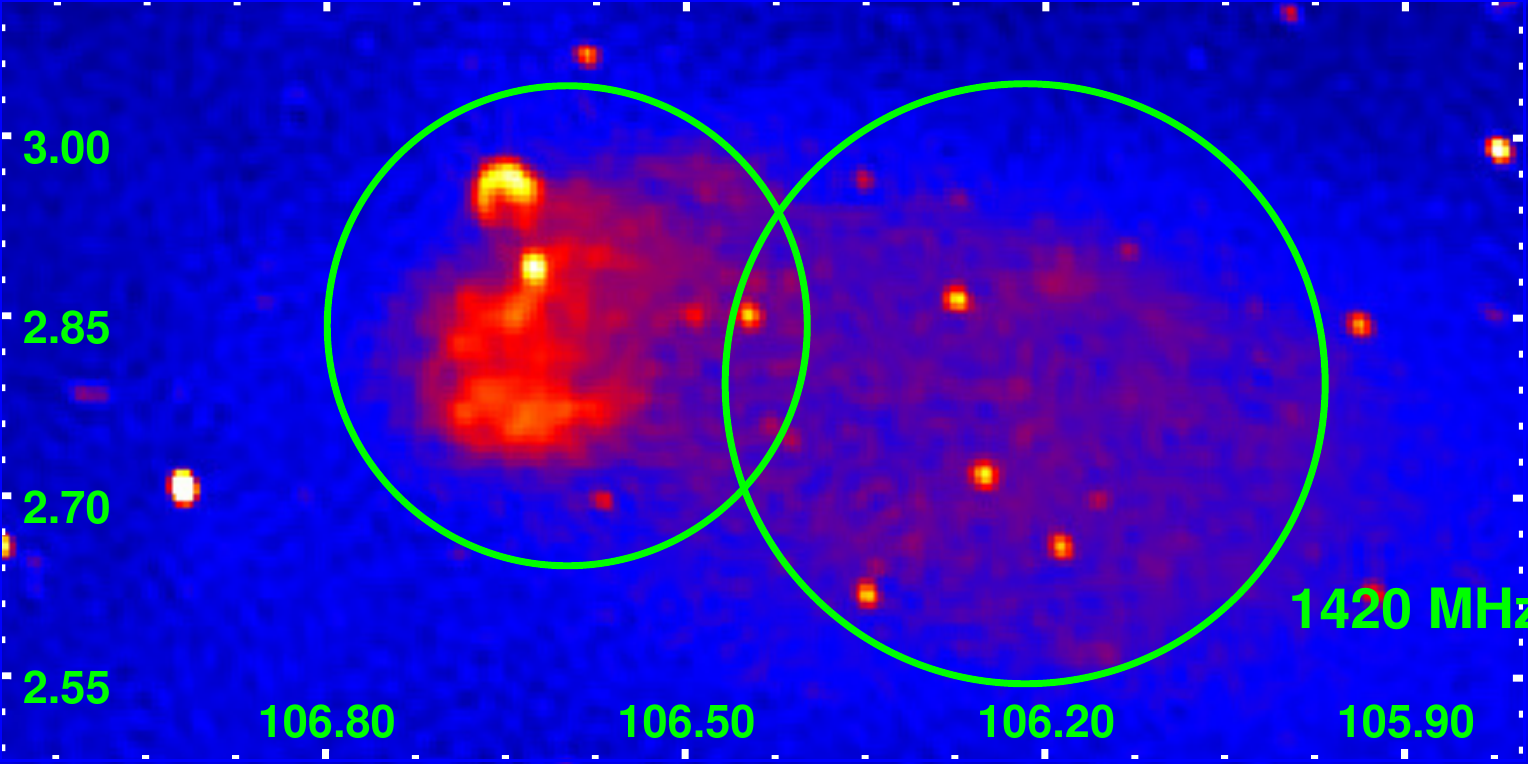}\includegraphics[width=0.3\textwidth, angle=0]{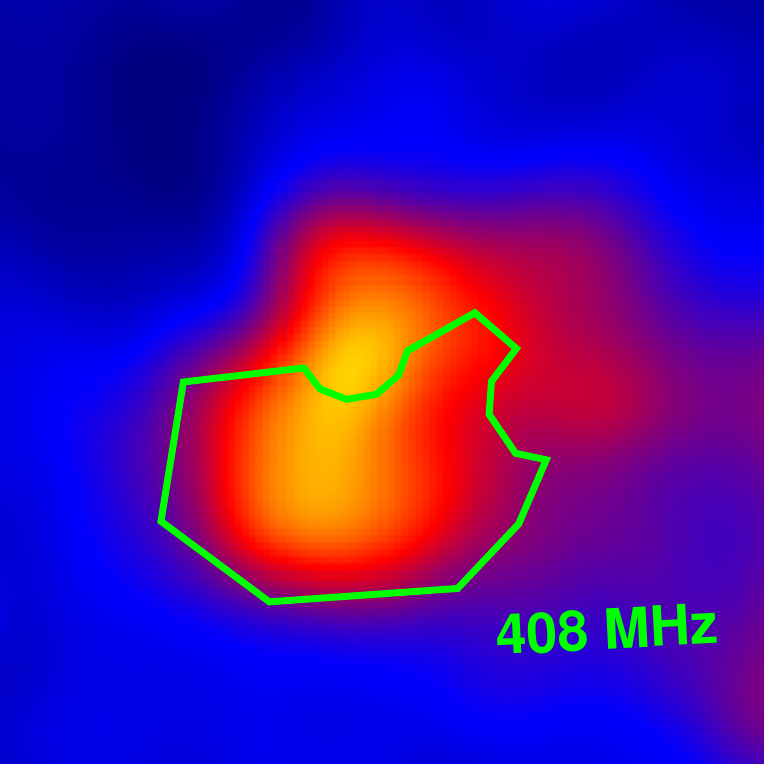}}
\caption{Mutli-wavelength images of the Boomerang PWN, the head and tail regions. The green lines in the top row indicate regions for radio flux extraction. The dashed and white lines in the first panel indicate the background regions for flux extraction of the head and PWN, respectively \citep{2006ApJ...638..225K}. The two black crosses mark two thermal point sources excluded from our flux extraction. The red contours in the second panel of the first row show the ASCA X-ray image \citep{2001ApJ...547..323H}. The small and large green circles in the first panel of the second row indicate the uniform templates used to extract GeV $\gamma$-ray fluxes from the head and tail regions, respectively.
\label{fig:0}}
\end{figure}

G106.3$+$2.7 was discovered by a 408 MHz radio survey having an elongated shape with a bright head to the northeast \citep{1990A&AS...82..113J}. It was later interpreted as an SNR with an unusual central bright head-tail morphology \citep[see also bottom-left panel of Fig. \ref{fig:0}]{2000AJ....120.3218P}. A radio and X-ray pulsar PSR J2229$+$6114 was identified at the north edge of the bright radio head having a spin-down luminosity of $2.2\times 10^{37}$ergs s$^{-1}$ and a characteristic age of $\sim 10$ kilo-years \citep{2001ApJ...552L.125H}. Pulsed $\gamma$-ray emission was also revealed with Fermi observations with a luminosity of $1.8\times 10^{34}$erg s$^{-1}$ \citep{2009ApJ...706.1331A} at a distance of 0.8 kpc determined via analyses of atomic hydrogen and molecular cloud distribution in the direction \citep{2001ApJ...560..236K}. This pulsar produces a prominent wind nebula, the so-called ``Boomerang'' PWN  with strongly polarized radio emission surrounding an extended diffuse X-ray source \citep[see also top-middle panel of Fig. \ref{fig:0}]{2001ApJ...547..323H}. With observations by {\em ROSAT} and {\em ASCA}, the total 2-10 keV flux of the extended diffuse X-ray source and the pulsar is found to be $1.56\times 10^{-12}$erg cm$^{-2}$ s$^{-1}$, with a power-law photon index of 1.51 $\pm$ 0.14 \citep{2001ApJ...547..323H}. With {\it Chandra} observations, \citet{2008ApJ...682.1166L} found that the flux from the pulsar is comparable to that of a compact X-ray PWN \citep{2008AIPC..983..171K}. The X-ray flux adopted here therefore is dominated by the extended source.

Radio observations show that the overall spectral index of the whole SNR is $\sim$ -0.57, and the tail region has a slightly steeper spectrum than the head region \citep{2000AJ....120.3218P}.
\citet{2011A&A...529A.159G} made the first observation of the entire SNR at $\lambda =$ 6 cm and obtained an integrated spectral index of $\sim -0.64$ with flux densities at $\lambda =$ 11 and 21 cm. In combination with the flux density at 408 MHz \citep{2006A&A...457.1081K}, there appears to be a spectral break near 500 MHz (Figure \ref{fig:1}).
For the ``Boomerang'' PWN, \citet{2006ApJ...638..225K} performed a multi-frequency analysis using observations at high radio frequencies by Effelsberg 100 m radio telescope and low radio frequencies from
the Canadian Galactic Plane Survey, and found that its radio continuum spectrum has a break between 4 and 5 GHz with the spectral index varying from $-0.11$ to $-0.59$ \citep{2006ApJ...638..225K}.
We re-analyzed these radio data and extracted flux densities from bright regions indicated by green lines in the top row of Figure \ref{fig:0}. The data points are shown as lower radio points in Figure \ref{fig:1}.

Very high-energy emission from the direction of G106.3$+$2.7 was first reported by the Milagro experiment \citep{2007ApJ...664L..91A}, which was later named VER J2227+608 by the VERITAS collaboration \citep{2009ApJ...703L...6A}. \citet{2019ApJ...885..162X} found prominent extended GeV emission with a spectral index of $1.81\pm 0.16$ from the tail region of G106.3$+$2.7. Here, besides a point source at PSR J2229$+$6114 and a uniform disk with a radius of 0.25 centered at (R.A., decl.)=($336.68^{\circ}$, $60.88^{\circ}$) for the Fermi source, we adopted an uniform disk with a radius of 0.2$^{\circ}$ centered at (R.A., decl.)=($337.30^{\circ}$, $61.12^{\circ}$) to calculate upper limits of GeV $\gamma$-ray emission from the head region. The results are shown as green arrows in Figure \ref{fig:1}. Recently, HAWC found that the TeV spectrum of VER J2227$+$608 can be fitted with a power-law with a spectral index of $2.29\pm 0.17$, and all these high-energy emission appears to be from the tail region of the SNR.

\section{Modeling}
\label{Mod}

Based on detection of a radio spectral break near $\sim 5$ GHz from the PWN, \citet{2006ApJ...638..225K} argued that the break is caused by synchrotron radiative energy loss of emitting electrons in a magnetic field of $\sim 3$ mG.  We find several caveats in this scenario. 1) In the relative compact region of $\sim 0.4$ pc, the magnetic field has a total energy of $\sim 10^{48}$ ergs. The volume of the whole SNR is at least 1000 times larger than that of the PWN. For a comparable magnetic field in the SNR, the magnetic field will have an energy greater than $10^{51}$ ergs, which is too high for a typical SNR \citep{2011ApJ...740L..26C}. Since the overall radio spectrum of G106.3$+$2.7 shows evidence of softening toward high frequencies \citep{2011A&A...529A.159G}, its spectrum may also has a break near $500$ MHz. If these spectral breaks have a common origin, the magnetic field in the SNR is then at most 10 times weaker than that in the PWN and is still too high for typical SNRs and PWNe.
2) For relativistic electrons accelerating in a magnetic field B, the synchrotron emission has a characteristic frequency of $\sim 4 B \gamma^2$MHz, where $\gamma$ is the electron Lorentz factor. Electrons producing the 5 GHz radio emission therefore have a Lorentz factor of $\sim 700$, which appears to be too low for particle acceleration in PWN. Given the hard spectrum below the break, this spectral feature is likely associated with the injection process \citep{2019MNRAS.490.3608O}. It is interesting to note that based on modeling of the structure of PWNe affected by reverse shocks of SNRs, \citet{2011ApJ...740L..26C} reached similar conclusions on the magnetic field and origin of the radio spectral break.

Here we explore the possibility that magnetic fields in the $\gamma$-ray emission region are so weak that we may attribute the $\gamma$-ray spectral break to energy loss of energetic electrons in the leptonic scenario for the $\gamma$-ray emission. Given the relatively weak X-ray emission, there is compiling evidence that magnetic fields in many extended TeV sources are less than 100 $\mu$G . In some extreme cases, the magnetic fields derived from multi-wavelength observations are within 10 $\mu$G \citep{2008PASJ...60S.163M,2011PASJ...63S.857F,2019A&A...627A.100H, 2019ApJ...875..149L}.

The processes of particle acceleration and transport in PWNe cover a large scale range and are complicated in general \citep{2019MNRAS.490.3608O}. We are mostly interested in large-scale features and acceleration processes at small scales may be described with stochastic particle acceleration in a turbulent medium \citep{1984A&A...136..227S}. In this context, we will adopt a quantitative model first proposed for hotspots in radio galaxies \citep{2008ApJ...673L.139F}. The model has two distinct zones. A steady acceleration region (Ac) is characterized by a particle injection rate, energy independent particle acceleration and escape times, whose ratio determines the index of particle escaping from the acceleration site, and a magnetic field, which in combination with the acceleration time determines the cutoff energy of electron distribution for the synchrotron radiative energy loss. Electrons escaping from the Ac are injected into an uniform cooling zone. In the absence of observational evidence for spectral hardening below the spectral cutoff, we will ignore the expansion of the Ac implying that the cooling zone has the same magnetic field as the Ac. There are therefore 6 model parameters: the injection rate and injection energy in the Ac, the acceleration and escape times, the magnetic field and age of the system.

\begin{figure}
\plottwo{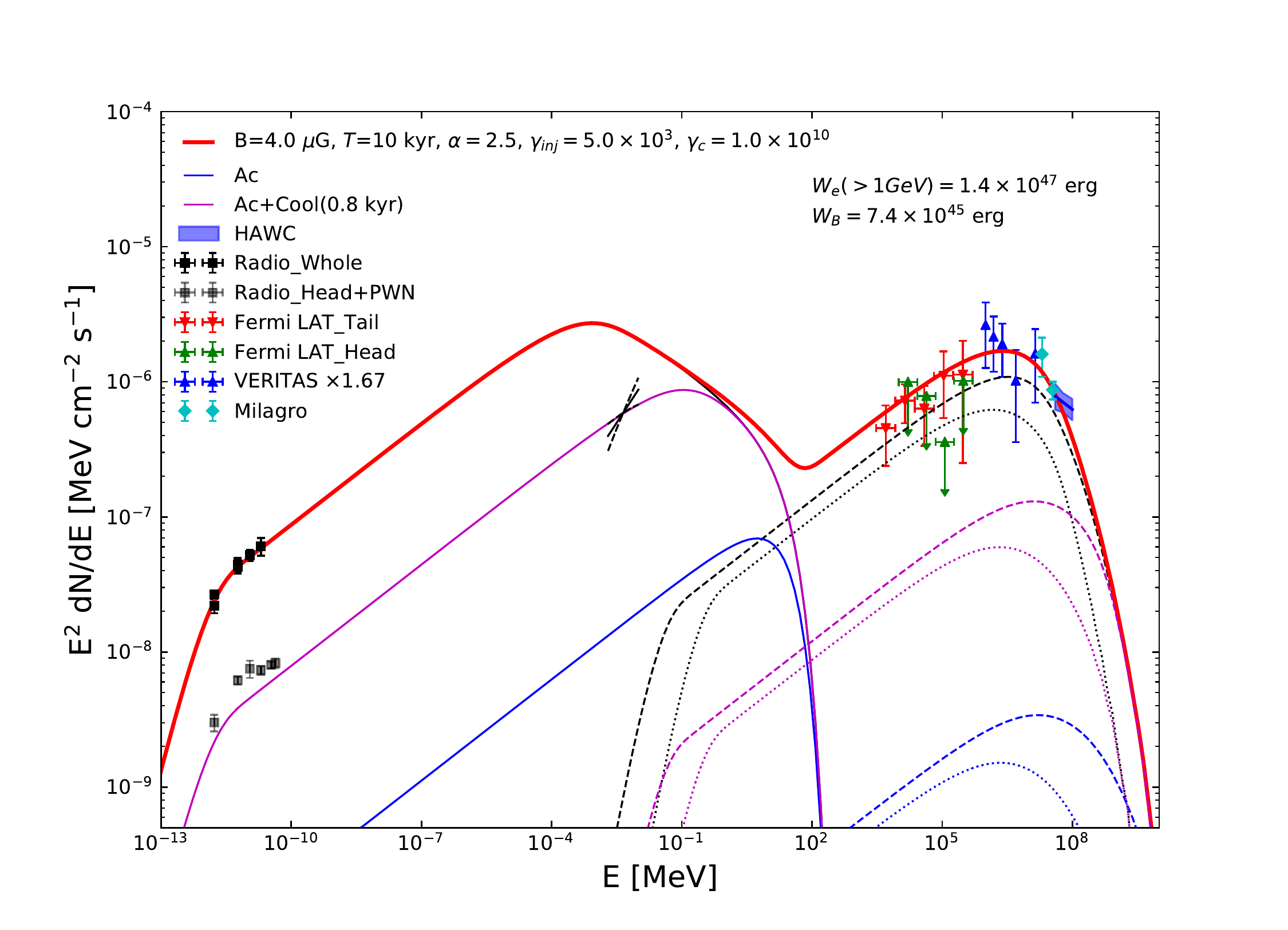}{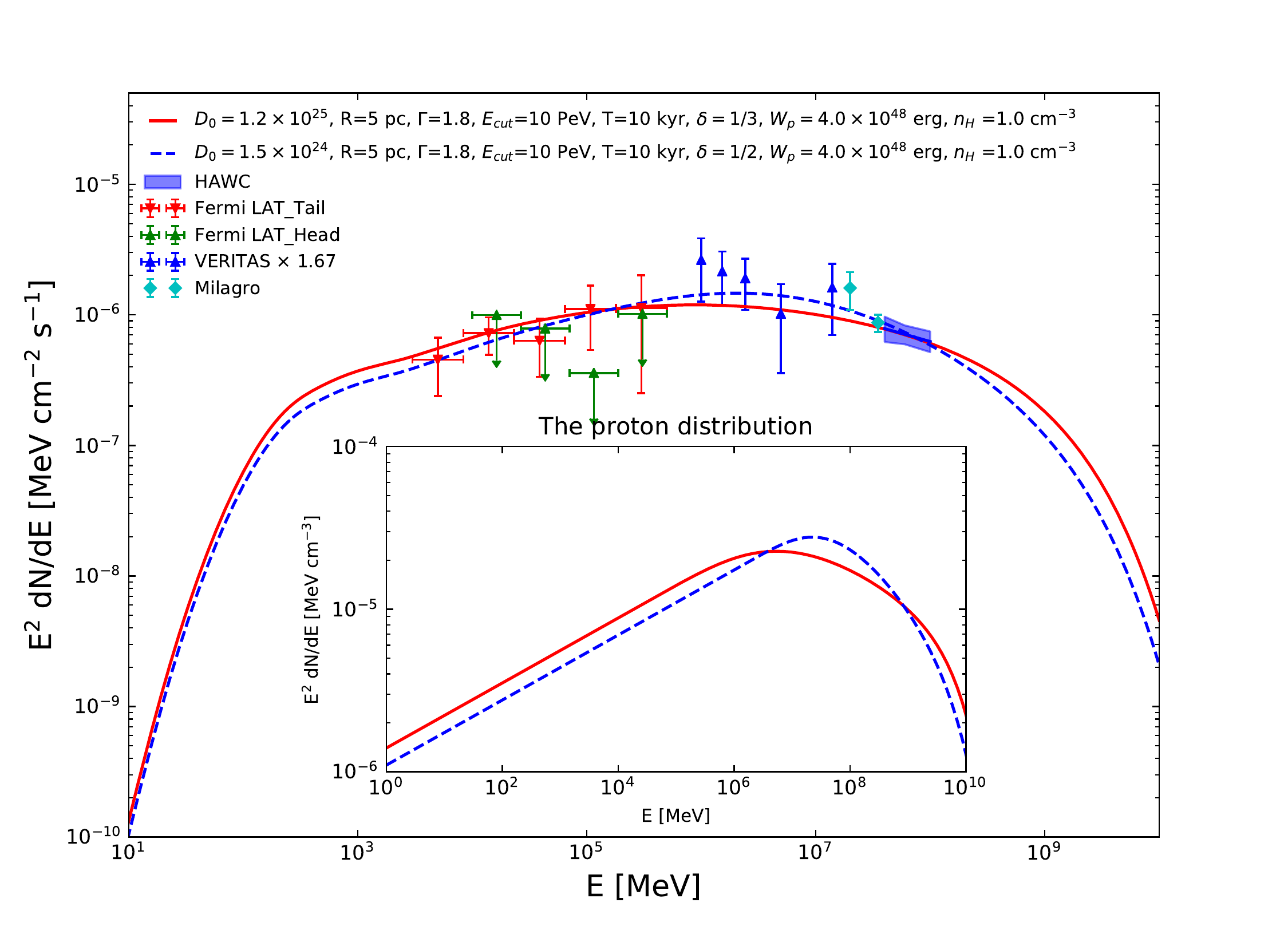}
\caption{Spectral fit to the SED of Ver J2227+608. Left: for the leptonic scenario. The thin solid, dashed, and dotted lines are for the synchrotron, IC of the CMB and NIR background, respectively. The blue lines correspond to emission from an acceleration region. The purple lines also include contributions from electrons escaping from the acceleration region over the past 800 years. The thin black lines are for all electrons. The thick solid line is for the overall emission spectrum. Right: $\gamma$-ray spectral fit for the hadronic scenario. The solid and dashed lines are for Kolmogorov and Kraichnan turbulences, respectively. Inserted is the averaged proton distribution in the emission region. Radio data of the whole region are from \cite{2000AJ....120.3218P,2006A&A...457.1081K,2011A&A...529A.159G}. The lower radio data are from this work.
The Fermi-LAT data are from \cite{2019ApJ...885..162X} and the upper limits are for the head region. The VERITAS data are from \cite{2009ApJ...703L...6A}, Milagro data from \cite{2007ApJ...664L..91A,2009ApJ...700L.127A}, and statistical uncertainties of HAWC from  \cite{2020arXiv200513699A}. The black butterfly of X-ray is from \cite{2001ApJ...547..323H}.
\label{fig:1}}
\end{figure}

The left panel of Figure \ref{fig:1} shows our fit to the multi-wavelength spectral energy distribution of G106.3+2.7 in the leptonic scenario for the $\gamma$-ray emission. For the inverse Compton (IC) emission, besides the cosmic microwave background radiation, we assume a near-infrared background with a temperature of 20 K and an energy density of 0.3 eV cm$^{-3}$ \citep{2017Sci...358..911A}.
The model parameters are indicated on the Figure. The magnetic field B $= 4   \,\mu$G and the age of the system T is fixed at 10 kyr. The electron distribution cuts off at $\gamma_{\rm c} = 10^{10}$. The acceleration time is given by the energy loss time at the cutoff energy $\gamma_{\rm c}{\rm m_e c}^2$: $\tau_{\rm ac} \simeq 25 \gamma_{\rm c}^{-1}B^{-2}$ years $\simeq 150$ years, which is longer than the gyro period of $\sim 28$ years for electrons with $\gamma_{\rm c}= 10^{10}$ moving in a magnetic field of $4 \,\mu$G.\footnote{Radiative energy loss due to IC processes has been ignored, which will affect the electron distribution near the cutoff energy slightly if the Klein-Nishina (KN) effect is important here. Otherwise, one may just use an effective magnetic field to rescale the acceleration time.} We note that $\gamma_c$ is not well constrained by our spectral fit and the value of $10^{10}$ can be considered as a lower limit to fit the HAWC spectral measurement. Theoretically, $\gamma_c$ can be a factor of 5 larger for the case of maximum acceleration in magnetic reconnection \citep{2011Sci...331..736T}.

The spectral index of electrons escaping the Ac is
$\alpha= 2.5$. The escape time is then $\tau_{\rm esc} = 4\tau_{\rm ac}/27\simeq 22$ years. The injection energy at the Ac is $\gamma_{\rm inj}=5\times 10^3$, which is constrained by the radio spectral break near 500 MHz. The injection rate is proportional to the total energy of emitting electrons. Here we give the total energy of electrons above $1$ GeV $W_{\rm e}= 1.4\times 10^{47}$ erg. For comparison, the total energy of uniformly distributed magnetic field in the SNR is about $7.4\times 10^{45}$ erg. These values are compatible with the leptonic model proposed by \citet{2019ApJ...885..162X}. Since we have a softer electron distribution, implying more electrons with relatively lower energies, the overall magnetic field is weaker than the $8 \mu$G field they found and the total energy of electrons above 1 GeV is higher than their value of $4\times 10^{46}$ erg.

The electron distribution in the cooling region has a broken power-law form with a break Lorentz factor $\gamma_{\rm b} \simeq 1.5\times 10^8$, where the radiative energy loss time is about 10 kyrs for the age of the SNR. Via the IC, such a distribution leads to a $\gamma$-ray spectral softening above a few TeVs due to the KN effect. Although the $\gamma$-ray spectrum is soft near $100$ TeV, it is still compatible with HAWC observations. Better spectral measurement beyond 100 TeV will be able to test this model.

The blue lines in the left-panel of Figure \ref{fig:1} show emission from the Ac, which should be associated with an X-ray bright region near the pulsar \citep{2008AIPC..983..171K}. However, the magnetic field at small scales should be high. At the surface of the neutron star, the magnetic field can be as high as $2 \times 10^{12}$G \citep{2001ApJ...552L.125H}. As mentioned above, the particle acceleration and transport near the pulsar are complicated, electrons with a small pitch angle to the magnetic field are expected to escape from the acceleration region more easily \citep{2019MNRAS.490.3608O}. The magnetic field we derived from the spectral fit therefore should be interpreted as an effective magnetic field for electrons with small pitch angles moving in a much stronger magnetic field.

The model can also explain the diffuse X-ray emission at the PWN with electrons escaping from the Ac during the past 800 years, as indicated by the purple lines in the left-panel of Figure \ref{fig:1}. There are two sets of radio data. The upper one is for the overall emission and the lower one is for bright regions at the PWN and the head. The X-ray emission region of the PWN only covers part of the radio bright regions and should have lower radio fluxes. The solid red line is for the overall emission from the SNR. We therefore expect soft diffuse X-ray emission from the head and tail regions with the X-ray spectral index increasing with the distance from the pulsar. Similarly, higher energy $\gamma$-rays should originate from regions closer to the pulsar. The lack of GeV $\gamma$-ray emission from the head may be attributed to its stronger magnetic field or softer electron distribution as suggested by its relatively softer radio spectrum shown in Figure \ref{fig:1}.

The hadronic scenario is relatively simple. We assume instantaneous injection of protons into an uniform emission zone with a density of n and a radius of R at T$=10$ kyrs ago. The injected protons have a power-law energy distribution with an exponential cutoff $E_{\rm cut}$:
$$
Q(E) = Q_0 E^{-\Gamma}\exp(-E/E_{\rm cut})\,,
$$
where $\Gamma$ is the spectral index.
We also assume a constant diffusion coefficient within R:
$$
D(E) = D_0 \left(\frac{E}{10{\rm GeV}}\right)^\delta\,.
$$
Then the proton distribution within the emission zone is given by
$$
N_{\rm p}(E, T) = Q(E) 4\pi \int_0^R r^2{\rm d}r  {1\over [4\pi D(E) T]^{3/2}}\exp\left[-{r^2\over 4D(E) T}\right].
$$
It is evident that $N_p$ only depends on $Q(E)$ and $R^2/D(E)T$.
We consider two values of $\delta$ for the diffusion coefficient. In a Kolmogorov turbulence, $\delta = 1/3$, while in the Kraichnan turbulence, $\delta=1/2$. The right panel of Figure \ref{fig:1} shows our fit to the $\gamma$-ray spectrum with $\Gamma=1.8$, $E_{\rm cut}=10$ PeV, and $R^2/D_0 T = 500$, and $63$ for $\delta=1/2$, and $1/3$, respectively. At low energies with $R^2\gg 4D(E)T$, $N_{\rm p}$ approximately equals to $Q(E)$. $N_p$ will deviate significantly (by a factor of $\sim 5$) from the power-law extension from low energies at $E_{\rm d}$ where $R^2=4D(E_{\rm d})T$. Then we have $E_{\rm d} = 160 \,,\ {\rm and}\ 39$ TeV for $\delta=1/2$, and $1/3$, respectively.
We note that the diffusion coefficients we have here are about one order of magnitude lower than mean values around TeV PWNe \citep{2020PhRvD.101j3035D} but are comparable to that derived for the Vela X \citep{2019ApJ...877...54B}. To increase the diffusion coefficient, one needs to increase the size of the emission zone or decrease the age T of the system so that the $\gamma$-ray spectrum does not change significantly.
The total energy of protons above 1 GeV is about $4\times 10^{48}(n/1{\rm cm}^{-3})^{-1}$ erg in the emission zone, which is consistent with that obtained by \citet{2019ApJ...885..162X}. The averaged energy density is about tens of eV/cm$^3$, which is typical for TeV halos \citep{2020A&A...636A.113G}. The corresponding total energy of injected protons is 8.5 and 11 $\times 10^{48}(n/1{\rm cm}^{-3})^{-1}$ erg for $\delta=1/2$, and $1/3$, respectively. Compared with the leptonic scenario, the hadronic models predict  much harder spectra at 100 TeV and beyond and an increase of the source extension with energy, which can be uncovered with future observations. Better $\gamma$-ray spectral measurement can also be used to constrain the diffusion coefficient.

The model here assumes spherical symmetry and efficient particle acceleration at the birth of the pulsar \citep{2018MNRAS.478..926O}, while the source is highly asymmetric. Since the pulsar was likely born at its current location \citep{2001ApJ...560..236K}, there is a convective flow from the pulsar to the $\gamma$-ray emitting tail region. Such a convection will lead to an increase of distance between the pulsar and $\gamma$-ray emission region with the decrease of energy for the energy dependent diffusion, reminiscent of the leptonic scenario. The overall $\gamma$-ray spectrum, however, will not be affected as far as the emission zone is more or less uniform.

\section{Conclusions}
\label{con}

We have shown that the broken power-law $\gamma$-ray spectrum of VER J2227+608 can be explained with both the leptonic and hadronic scenarios for the $\gamma$-ray emission with reasonable parameters. These two models, however, predict distinct spectra above 100 TeV. High resolution observations such as LHAASO and CTA should be able to test these models.

The hadronic model attributes the spectral break to an energy dependent diffusion process. To account for the observed $\gamma$-ray flux, the total energy of relativistic protons injected into the emission region is about $10^{49}(n/1{\rm cm}^{-3})^{-1}$ erg. Although it is tempting to attribute to such a population of protons to shock acceleration in the early stage of SNR evolution, as the only SNR with emission above 100 TeV, its pulsar may also play an important role in the proton acceleration \citep{2018MNRAS.478..926O}. Other powerful young PWNe then should have a similar emission component if there are sufficient background materials to support efficient hadronic emission. The model also requires a very low diffusion coefficient, commonly revealed in extended TeV sources \citep{2020PhRvD.101j3035D}.

The leptonic scenario is very interesting in light of extensive multi-wavelength studies of Vela X PWN and its pulsar, which has a similar characteristic age and a 4 times lower spin down luminosity \citep{2011ApJ...740L..26C}. The center of the TeV source is also displaced from the pulsar. The TeV spectrum of Vela X cuts off at a bit above 10 TeV, which has been attributed to radiative energy loss of TeV electrons displaced by reverse shocks of the SNR a few thousand years ago \citep{2019ApJ...877...54B}. In our model, we have a constant injection rate of high-energy electrons into the cooling region, implying a very long spin down age for PSR J2229+6114.  It is possible that the spin down age of Vela pulsar is within a few thousand years so that injection over the past few thousand years are negligible \citep{2019ApJ...875..149L}. Nevertheless, we expect some flux of $\gamma$-ray beyond 100 TeV from Vela X due to recent injection of high-energy electrons into the nebula. Detailed modeling of such kind of PWNe are warranted \citep{2011ApJ...740L..26C}.

\acknowledgments

We would like to thank the anonymous referee for very helpful comments, which help to improve the paper.
We thank Roland Kothes for providing the radio images initially presented in \cite{2006ApJ...638..225K}. This work is partially supported by the National Key
R\&D Program of China 2018YFA0404203, NSFC grants U1738122, U1931204, U1738125, U1938112 and 11761131007, the Natural Science Foundation for Young Scholars of Jiangsu Province, China (No. BK20191109), and by the International Partnership Program of Chinese Academy of Sciences, grant No.114332KYSB20170008.

%





\end{document}